\documentclass[epjCONF]{svjour}
\usepackage{graphics}
\usepackage{upgreek}
\usepackage{multicol}
\usepackage{amsmath}
\usepackage[varg]{txfonts} 
\usepackage[latin1]{inputenc}
\session-title{Probing stellar structure and evolution with asteroseismology}
\begin{document}
\title{\'Echelle diagrams and period spacings of g modes in
  $\boldsymbol{\gamma}$~Doradus stars from four years of {\em Kepler}
  observations}

\author{Timothy
R. Bedding\inst{1,2}\fnmsep\thanks{\email{t.bedding@physics.usyd.edu.au}}
\and Simon J. Murphy\inst{1,2} \and Isabel L. Colman\inst{1} \and Donald
W. Kurtz\inst{3}}
\institute{Sydney Institute for Astronomy (SIfA), School of Physics,
  University of Sydney, Australia \and Stellar Astrophysics Centre,
  Department of Physics and Astronomy, Aarhus University, Denmark \and
  Jeremiah Horrocks Institute, University of Central Lancashire, Preston,
  PR1 2HE, UK }
\abstract{We use photometry from the {\em Kepler} Mission to study
  oscillations in $\gamma$~Doradus stars.  Some stars show remarkably clear
  sequences of g~modes and we use period \'echelle diagrams to measure
  period spacings and identify rotationally split multiplets with $\ell=1$
  and $\ell=2$.  We find small deviations from regular period spacings that
  arise from the gradient in the chemical composition just outside the
  convective core.  We also find stars for which the period spacing shows a
  strong linear trend as a function of period, consistent with relatively
  rapid rotation.  Overall, the results indicate it will be possible to
  apply asteroseismology to a range of $\gamma$~Dor stars. }

\maketitle
%
\label{intro}

Gravity modes are extremely valuable for probing stellar interiors.
Asteroseismology using g~modes has so far produced very good results on
three classes of highly evolved stars, namely white dwarfs~\cite{W+K2008},
sdB stars~\cite{kepler-reed++2011-compact-VIII} and red
giants~\cite{kepler-bedding++2011-rg-nature,kepler-beck++2012-rg-rotation,kepler-mosser++2012-rg-rotation}.
Excellent results have also been obtained for a few SPB stars (slowly
pulsating B stars), which lie on the upper main
sequence~\cite{corot-degroote++2010-hd50230-nature,papics++2012-hd43317}.
However, the g modes lower on the main sequence, which occur in
$\gamma$\,Doradus stars, have proved much more difficult to exploit.  They
typically have periods close to one day, which makes ground-based study
exceedingly difficult.  Furthermore, as we show here, they have very dense
frequency spectra.  Even the first month or so of \textit{Kepler} data,
which revealed many stars with g modes (and many hybrids having both g and
p modes), was not enough to properly resolve their frequency
spectra~\cite{kepler-grigahcene++2010-hybrids}.  With four years of
nearly-continuous photometry from {\em Kepler}, we are finally in a good
position to apply asteroseismology to $\gamma$\,Dor stars.

Applying asteroseismology requires identifying which modes are excited.  In
$\gamma$\,Dor stars, we are guided by the expectation that g modes should
be approximately equally spaced in period, at least for slow rotators.  So
far, only one $\gamma$\,Dor star has been reported with a clearly measured
period spacing.  This is KIC\,11145123, which was found to have a series of
rotationally split $\ell=1$ triplets with a regular period spacing of
$\Delta P = 2100$\,s~\cite{kurtzetal2014}.  Here, we look at this star and
other $\gamma$~Dor pulsators in the {\em Kepler} field and show that some
of them have remarkably clear sequences of g~modes.


Solar-like stars have p-mode oscillations that are approximately equally
spaced in frequency.  The so-called \'echelle diagram is made by dividing
the frequency spectrum into equal segments and stacking them one above the
other so that modes with a given degree align vertically in
ridges~\cite{GFP83}.  Any departures from regularity are clearly visible as
curvature in the \'echelle diagram.  For g-modes, the regularity is in
period rather than frequency, which suggests the use of a period \'echelle
diagram~\cite{kepler-bedding++2011-rg-nature}.  Note that the period
spacing of g~modes decreases with angular degree according to $\Delta P
\propto 1/\sqrt{\ell(\ell+1)}$, which means a different \'echelle diagram
is needed for each value of~$\ell$.

\begin{figure}
\centering
\resizebox{0.60\columnwidth}{!}{%
\includegraphics{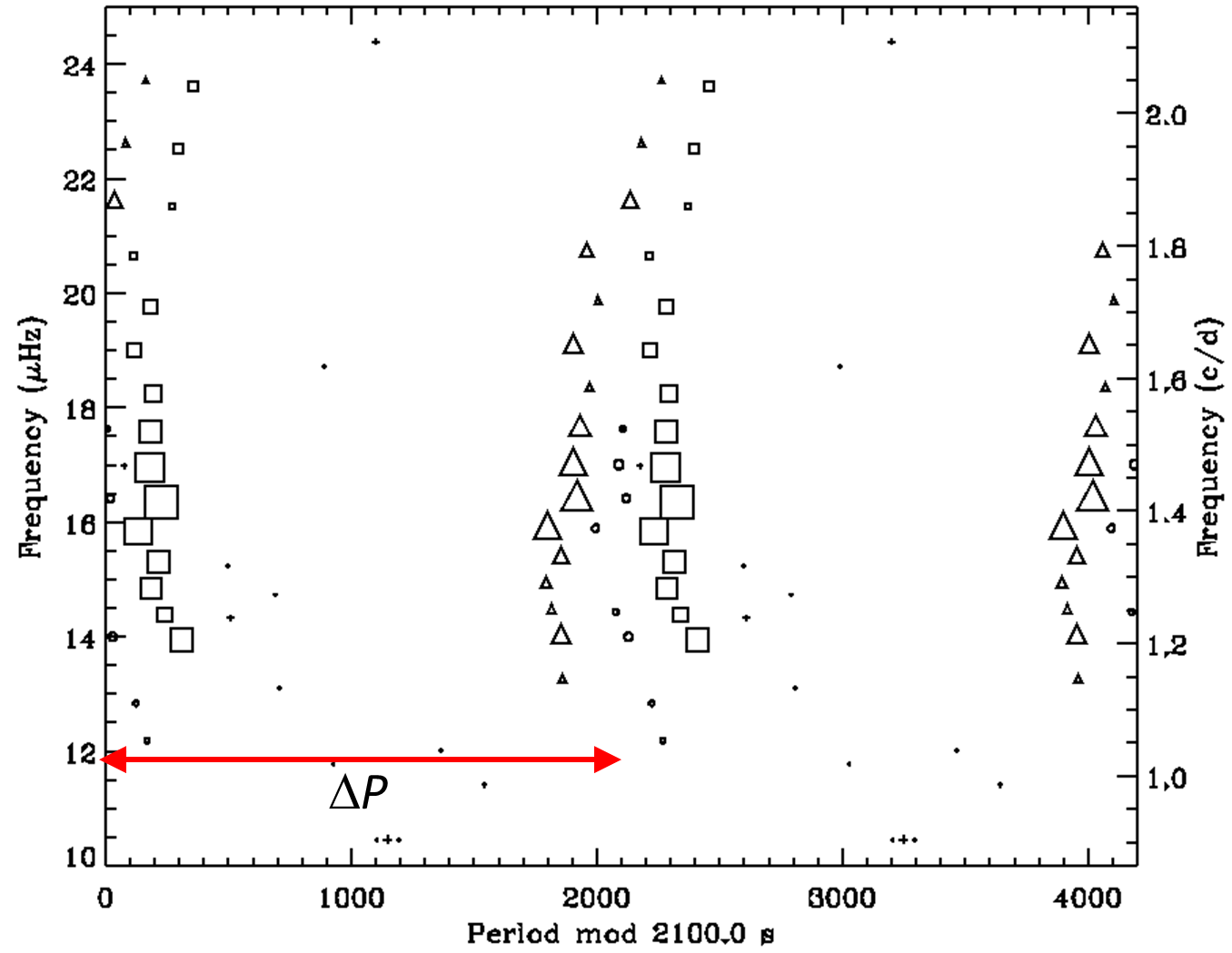} }
\hfill
\resizebox{0.36\columnwidth}{!}{%
\includegraphics{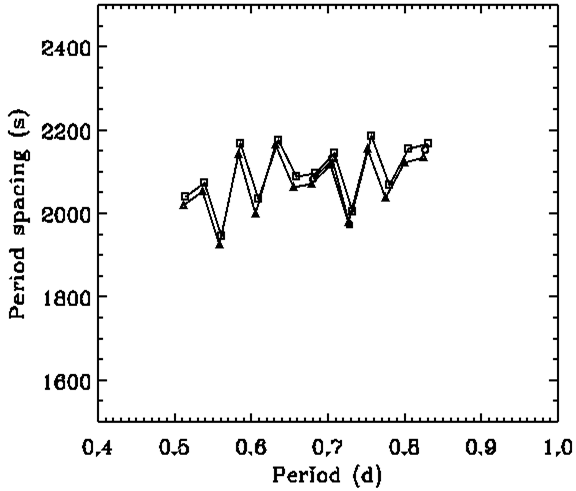} }
\caption{Period \'echelle diagram, plotted twice for clarity ({\em left}),
and period spacings ({\em right}) for KIC\,11145123.  Symbol sizes indicate
mode amplitudes and symbol shapes indicate different values of azimuthal
degree~$m$.}
\label{fig:11145123}    
\end{figure}

Figure\,1 (left panel) shows the period \'echelle diagram for
KIC\,11145123~\cite{kurtzetal2014}, plotted twice for clarity.  The peaks
shown in this diagram were extracted by iterative sine-wave fitting, also
known as ``CLEAN'' or ``prewhitening'', and the symbol size indicates mode
amplitude.  The diagram clearly shows a series of triplets, although $m=0$
is weak and is only visible in some orders.  We also see ``wiggles'', which
indicate small departures from regular spacing.  This is further
illustrated in the right panel of Fig.~\ref{fig:11145123}, which shows
pairwise differences between consecutive modes for $m=-1, 0$ and~$1$.
These irregularities were predicted theoretically~\cite{miglioetal2008} and
are caused by the gradient in the chemical composition just outside the
convective core.  This phenomenon is well studied in white dwarfs, sdBs and
SPBs, and is sometimes referred to as mode trapping.

\begin{figure}
\centering
\resizebox{0.60\columnwidth}{!}{%
\includegraphics{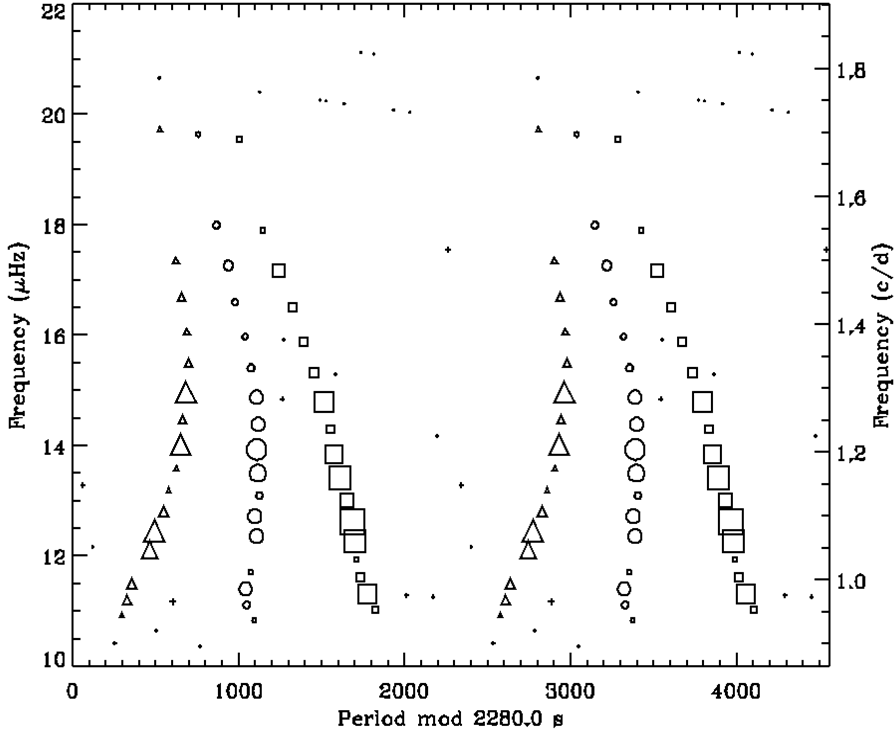} }
\hfill
\resizebox{0.36\columnwidth}{!}{%
\includegraphics{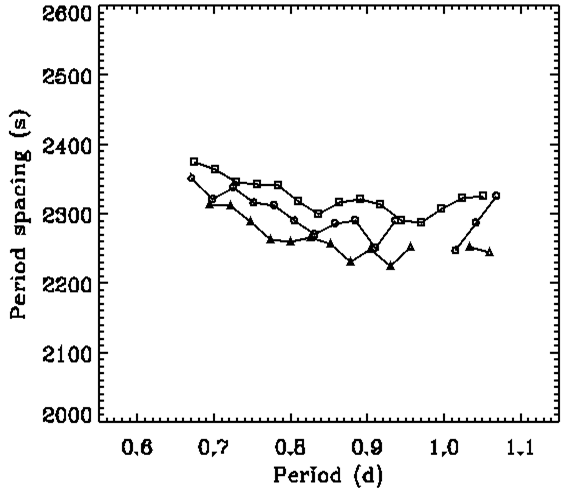} }
\caption{Same as Fig.~\ref{fig:11145123} but for KIC\,9244992.}
\label{fig:9244992}    
\end{figure}

A second example (KIC\,9244992) is shown in Fig.~\ref{fig:9244992} and we
see a clear sequence of $\ell=1$ triplets with small deviations in period
spacing.  As with many of the stars we have examined, almost all the
extracted peaks are identified and there are almost no modes missing.  This
makes these stars very nice for asteroseismology.

\begin{figure}
\centering
\resizebox{1.00\columnwidth}{!}{%
\includegraphics{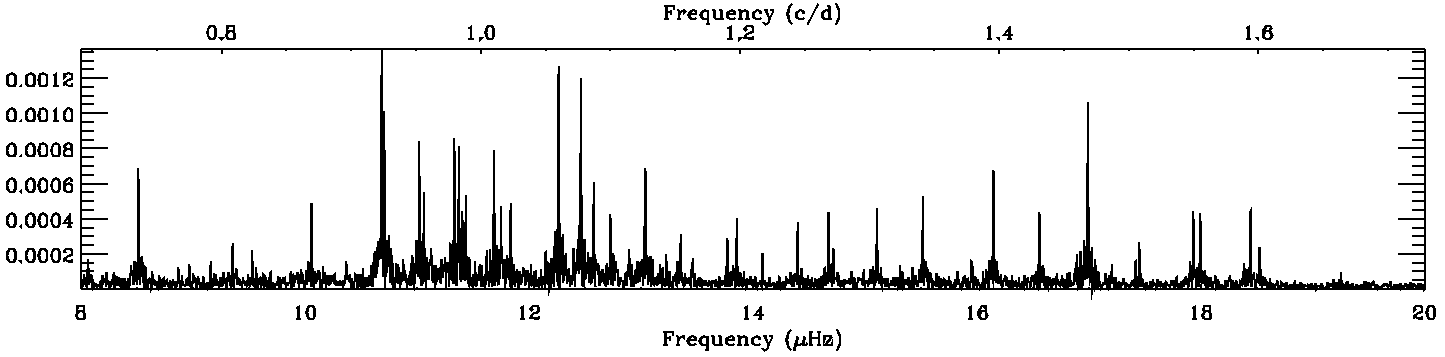} }

\smallskip

\resizebox{0.48\columnwidth}{!}{%
\includegraphics{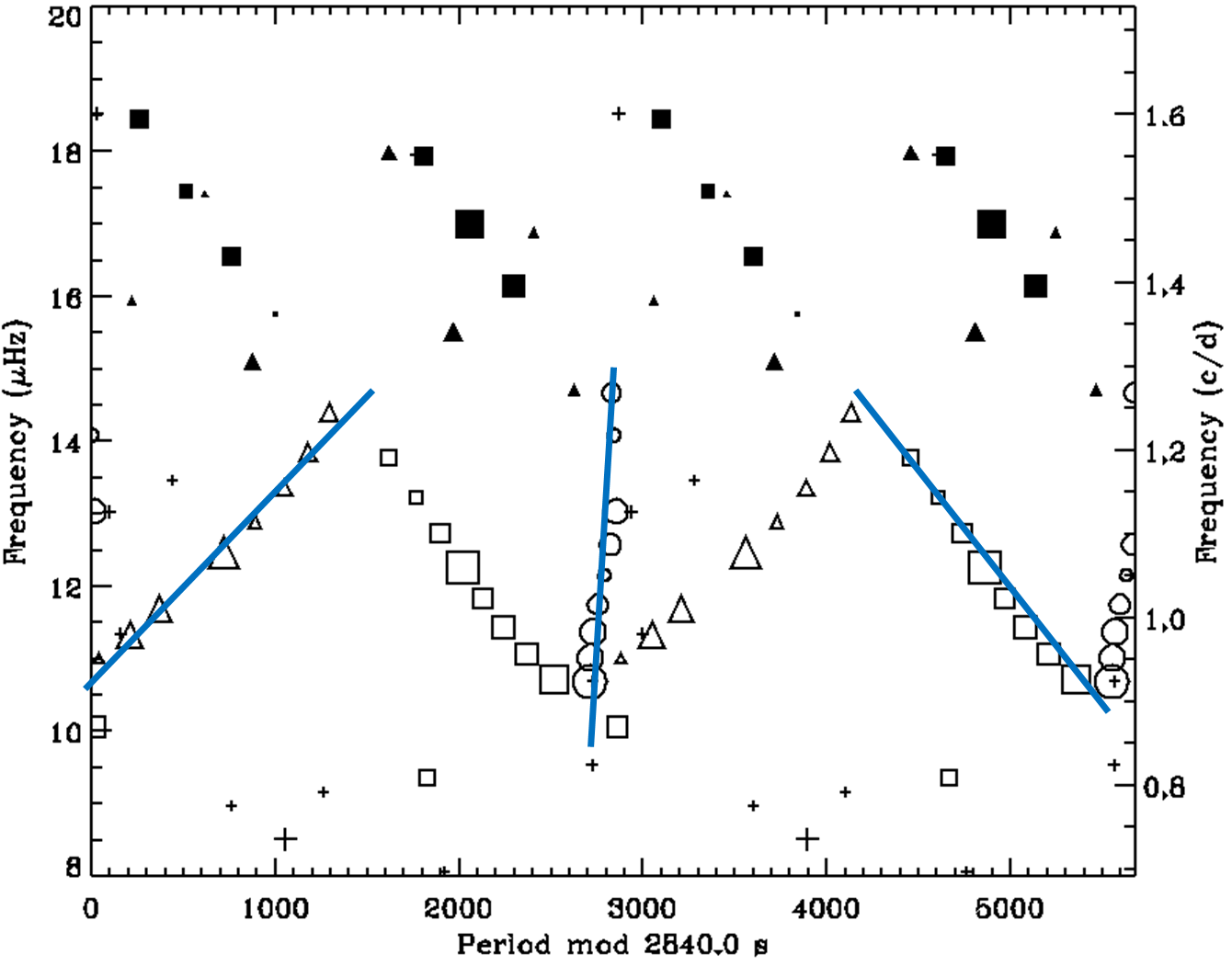} }
\hfill
\resizebox{0.47\columnwidth}{!}{%
\includegraphics{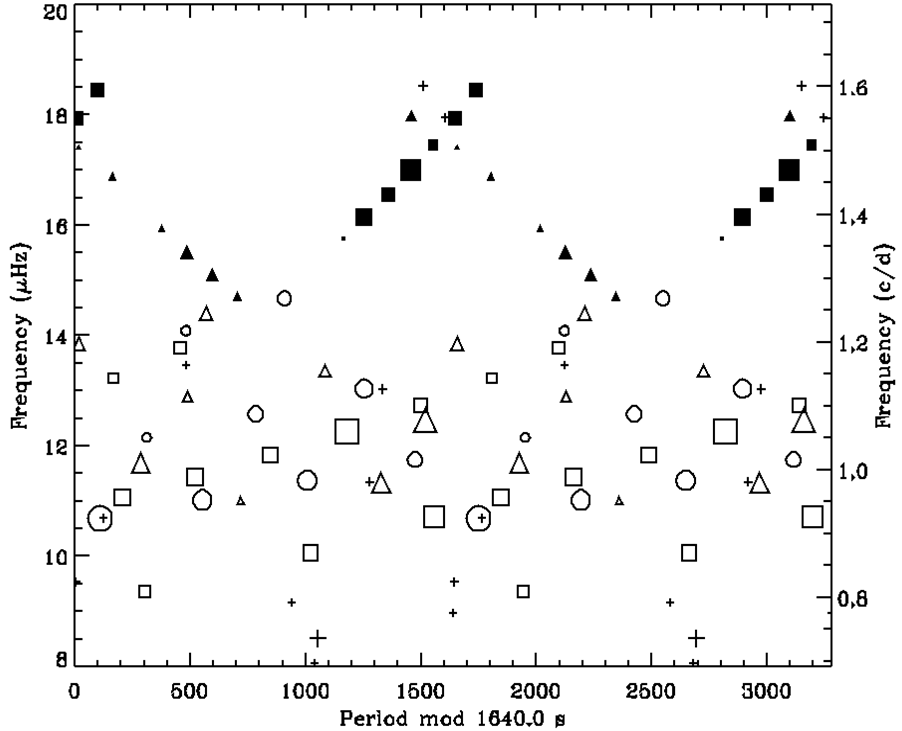} }

\smallskip

\resizebox{1.00\columnwidth}{!}{%
\includegraphics{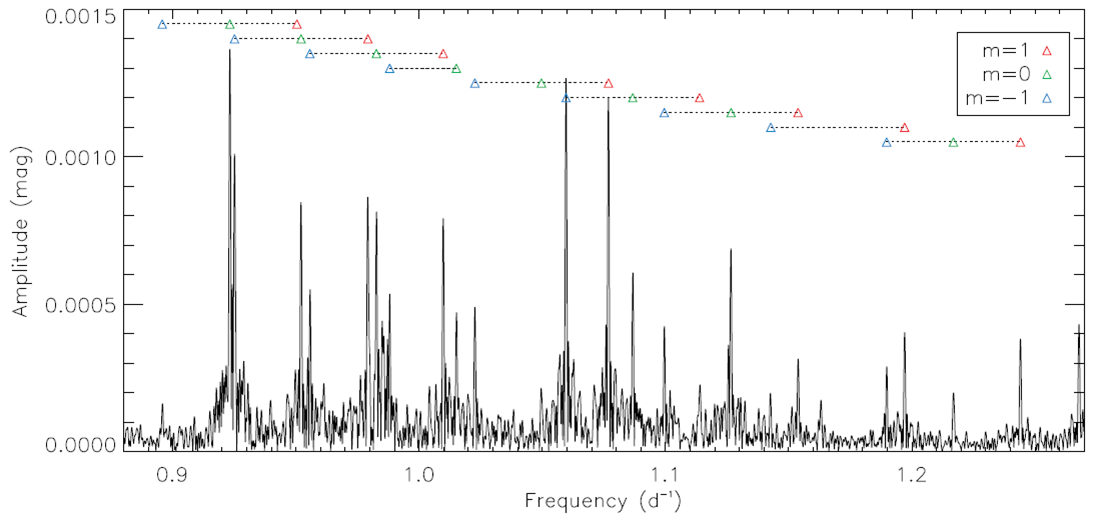} }
\caption{Analysis of KIC\,3127996.  {\em Top:} amplitude spectrum showing
  the region containing g~modes.  {\em Middle left:} period \'echelle
  diagram with a spacing of $\Delta P = 2840$\,s, showing the $\ell=1$
  modes (open symbols).  {\em Middle right:} period \'echelle using a
  spacing that is smaller by a factor of $\sqrt3$, showing the alignment of
  $\ell=2$ modes (filled symbols).  {\em Bottom:} the region of the amplitude
  spectrum containing the $\ell=1$ triplets, where rotational splitting has
  caused them to overlap.}
\label{fig:3127996}    
\end{figure}

Our third example (Fig.~\ref{fig:3127996}), is rotating more rapidly.  This
causes the triplets to overlap and the ridges in the \'echelle diagram are
spread more widely (blue lines).  Indeed, finding the correct period
spacing requires looking at histogram of pairwise differences and some
trial-and-error.  Note that the amplitude spectrum (top panel) contains two
humps of peaks.  Our \'echelle diagrams indicate that the higher-frequency
hump has a period spacing that is $\sqrt{3}$ times smaller than the
lower-frequency hump.  This implies that the two humps correspond to modes
with $\ell=1$ and $\ell=2$.

\begin{figure}
\centering
\resizebox{0.60\columnwidth}{!}{%
\includegraphics{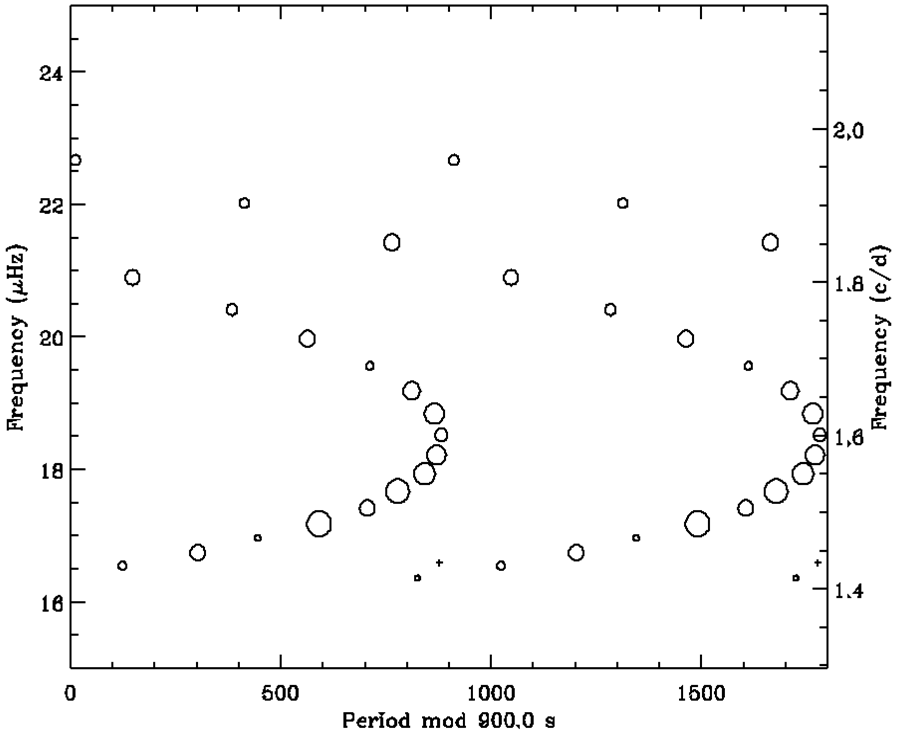} }
\resizebox{0.36\columnwidth}{!}{%
\includegraphics{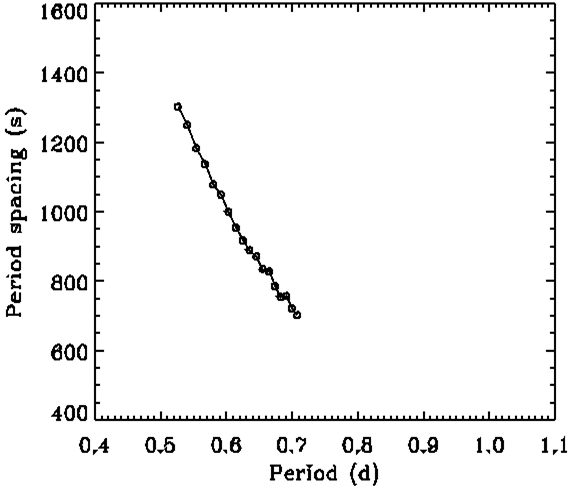} }
\caption{Period \'echelle diagram, plotted twice for clarity ({\em left}),
  and period spacings ({\em right}) for KIC\,4253413.} 
\label{fig:4253413}    
\end{figure}

\begin{figure}
\centering
\resizebox{0.60\columnwidth}{!}{%
\includegraphics{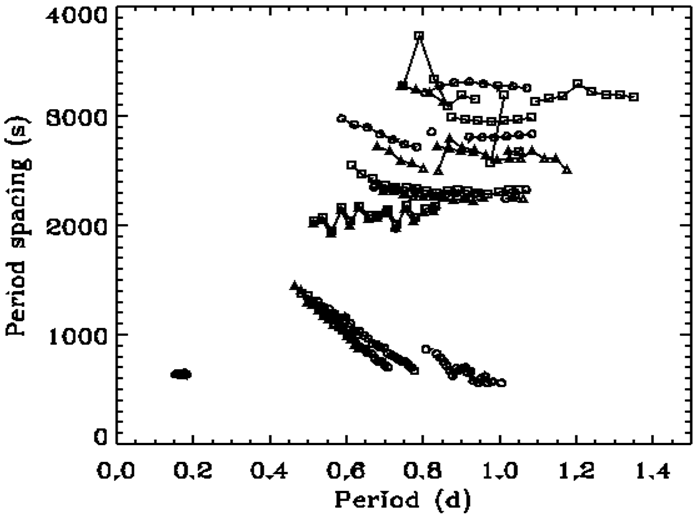} }
\caption{Period spacings of $\ell=1$ modes for twelve \textit{Kepler}
  $\gamma$\,Dor stars, including those shown above.  For each star, symbol
  shapes indicate different values of azimuthal degree~$m$. }
\label{fig:period-spacings}    
\end{figure}

Figure~\ref{fig:4253413} shows an \'echelle diagram with a single parabolic
ridge (left panel), indicating a period spacing that varies linearly with
period (right panel).  This is consistent with calculations of a rotating
star with mixing from diffusion (see Fig.\,7 of~\cite{bouabidetal2013}).
We have found quite a few more examples of this behaviour.
Figure~\ref{fig:period-spacings} combines the stars mentioned in this
paper, as well as several others that were shown in the conference talk,
including one evolved star having much faster pulsations (0.2\,d).
Overall, these results indicate the exciting possibility of applying
asteroseismology to a range of $\gamma$~Dor stars.

\begin{multicols}{2}

\end{multicols}

\end{document}